\documentclass{WileyMSP-template}
\usepackage{xcolor}
\usepackage{setspace}

\begin{document}


\title{Insights into the Nature of Quantum Emitters in Electron-Irradiated Hexagonal Boron Nitride}

\maketitle

\author{Mouli Hazra$^{1,3,*}$}
\author{Anna Rupp$^{2}$}
\author{Mohammad N.~Mishuk$^{1,3}$}
\author{Josefine Krause$^{1,3,4,6}$}
\author{Anand Kumar$^{1,3}$}
\author{Julien Chénedé$^{1,3}$}
\author{Mingi Kang$^{1,3}$}
\author{Bayarjargal N.~Tugchin$^{4}$}
\author{Marijn Rikers$^{4,7,8}$}
\author{Isabelle Staude$^{4,7}$}
\author{Thomas Pertsch$^{4,5,6}$}
\author{Alexander Högele$^{2,3}$}
\author{Tobias Vogl$^{1,3,*}$}

\dedication{}

\begin{affiliations}
$^{1}$ Department of Computer Engineering, TUM School of Computation, Information and Technology,  
Technical University of Munich, 80333 Munich, Germany

$^{2}$ Fakultät für Physik, Munich Quantum Center, and Center for NanoScience (CeNS),  
Ludwig-Maximilians-Universität München, 80539 Munich, Germany

$^{3}$ Munich Center for Quantum Science and Technology (MCQST),  80799 Munich, Germany

$^{4}$ Institute of Applied Physics, Abbe Center of Photonics,  
Friedrich Schiller University Jena, 07745 Jena, Germany

$^{5}$ Fraunhofer-Institute for Applied Optics and Precision Engineering IOF, 07745 Jena, Germany

$^{6}$ Max Planck School of Photonics, 07745 Jena, Germany

$^{7}$ Institute of Solid State Physics,  
Friedrich Schiller University Jena, 07743 Jena, Germany

$^{8}$ ARC Centre of Excellence for Transformative Meta-Optical Systems,  
Department of Quantum Science and Technology,  
Research School of Physics, Australian National University,  
Canberra, ACT 2601, Australia

\bigskip
$^{*}$ Corresponding author:  
\texttt{mouli.hazra@tum.de; tobias.vogl@tum.de}

\end{affiliations}



\begin{abstract}
Quantum emitters in hexagonal boron nitride (hBN) have emerged as a promising solid-state platform for quantum technology applications. However, a persistent challenge in the field is the unclear origin of many observed emission lines, particularly in the visible range, which can be difficult to distinguish from signals arising from organic or process-induced contamination during sample preparations and handling. This ambiguity limits both the reproducibility of emitter generation and the reliable identification of truly intrinsic quantum defects. This work provides a step-by-step framework to assess whether quantum emitters in electron-irradiated hBN are associated with organic contaminants introduced during sample preparation. We employ hyperspectral imaging, thermal annealing, and oxygen plasma etching to investigate the origin of the green-yellow emitters in electron-irradiated hBN. The combined results not only rule out organic contamination as the source of emission but also provide insight into the spectral variability, thermal stability, and vertical localization of the emitters generated in electron-irradiated hBN that was created without any pre- or post-processing. In addition, our experiments demonstrate the feasibility of creating stable emitters in hBN with thicknesses below 10~nm. These findings provide practical guidance for the identification and controlled implementation of hBN-based single-photon emitters in quantum photonic devices.

\end{abstract}


\section{Introduction}
\indent Hexagonal boron nitride (hBN) provides a distinctive set of material advantages in scalable quantum technologies. Its wide bandgap \cite{cassabois2016hexagonal} supports a diverse range of room temperature (RT) quantum emitters spanning from the near-infrared to the ultraviolet \cite{shevitski2019blue,cholsuk2022tailoring,cholsuk2024hbn,camphausen2020observation, bourrellier2016bright}. Its two-dimensional, atomically flat structure enables precise spatial control and seamless integration into on-chip photonic architectures \cite{li2021integration,zalogina2026engineering,mishuk2025all,gerard2023top}. This planar geometry not only simplifies fabrication processes but also improves optical accessibility, enhancing both excitation efficiency and photon collection \cite{fröch2021coupling, gerard2024quantum}. Beyond fundamental quantum physics experiments \cite{vogl2021sensitive}, demonstrated applications with hBN include sensing of temperature \cite{hazra2026temperature,akbari2021temperature, ari2025temperature}, pressure \cite{gottscholl2021spin}, strain \cite{grosso2017tunable}, electric \cite{zhigulin2023stark} and magnetic fields \cite{gottscholl2021spin,stern2022room,  mu2022excited}, as well as quantum cryptography \cite{samaner2022free}, quantum computing \cite{conlon2023approaching}, quantum random number generation \cite{hoese2022single} and space instrumentation \cite{ahmadi2024quick,vogl2019radiation}.\\
\indent Despite these advancements, the practical deployment of hBN-based quantum emitters remains limited by an incomplete understanding of their microscopic origin and the degree of control over their formation. These challenges are closely interconnected, as precise and reproducible emitter generation requires detailed knowledge of the underlying defect structures and their interaction with the local environment. To date, hBN emitters have exhibited substantial variability in their spectral characteristics \cite{grosso2017tunable,tran2016robust,shotan2016photoinduced,mendelson2019engineering,martinez2016efficient} and lifetime \cite{cholsuk2024hbn,grosso2017tunable,tran2016robust,jungwirth2017optical,vogl2017room,vogl2019atomic,kumar2024polarization,tran2016quantum,exarhos2017optical,schell2018quantum}, with emission properties strongly influenced by strain \cite{mendelson2019engineering,aharonovich2016solid}, temperature \cite{hazra2026temperature,akbari2021temperature,ari2025temperature,jungwirth2017optical,du2014temperature}, electric \cite{zhigulin2023stark,dhu2024electrical}  and magnetic fields \cite{gottscholl2021spin,stern2022room,gottscholl2021room}, and the surrounding dielectric environment \cite{gerard2024quantum,yamamura2025quantum}. While this pronounced environmental sensitivity makes these emitters promising candidates for nanoscale sensing applications, it poses a challenge for quantum photonic technologies, where stable and reproducible emission is required. This variability has raised fundamental questions about the intrinsic versus extrinsic nature of the observed emission. In particular, several observations suggest that certain optical signatures attributed to hBN defects may instead originate from extrinsic sources, such as organic contaminants \cite{neumann2023organic}. Several classes of organic molecules are known to exhibit photophysical properties remarkably similar to those reported for single photon emitters (SPEs) in hBN, including visible-range emission, nanosecond-scale lifetimes, pronounced vibronic sidebands, and high single-photon purity \cite{ neumann2023organic,neumann2019accidental,zhao2018single, chen2019dibenzo,macklin1996imaging,qazi2021detection}. Given the wide bandgap of hBN \cite{cassabois2016hexagonal}, which can suppress non-radiative quenching, the material can act as an efficient host matrix for such molecular emitters \cite{shkarin2026organic}. This has led to ongoing debate as to whether at least a subset of reported emitters arises from contamination rather than intrinsic lattice defects.\\
\indent Yellow quantum emitters in electron-irradiated hBN have been demonstrated to exhibit reproducible and stable single-photon emission, with a characteristic emission peak around 575~nm at RT and high single-photon purity \cite{mishuk2025all,kumar2024polarization,kumar2023localized}. At cryogenic temperatures, a well-resolved zero phonon line (ZPL) emerges at ~547.5~nm, indicating that these so-called “yellow” emitters more precisely fall within the green–yellow spectral range \cite{hazra2026temperature}. However, as is the case for most hBN emitters reported so far, their microscopic origin remains elusive. In addition, low-acceleration-voltage electron-beam irradiation is known to induce the deposition of a carbonaceous film originating from residual hydrocarbons on the sample surface and/or scanning electron microscope (SEM) chamber \cite{postek1996approach,nedic2024electron}. There are several unavoidable sources of organic contamination during sample preparation. For instance, mechanical exfoliation using adhesive tape can leave glue residues; polymers are commonly used to transfer hBN flakes onto substrates. No pre- or post-processing steps such as annealing, plasma or UV cleaning were performed in this case. While minimizing processing steps is advantageous in terms of simplicity and time efficiency, it also increases the likelihood of introducing emitters that are not intrinsic to the hBN lattice. This motivated a systematic investigation of the emitter origin. Rather than modifying the fabrication process, we performed a step-by-step analysis of the emission properties using a combination of experimental techniques. \\
\indent Firstly, we performed hyperspectral mapping on electron-irradiated hBN to investigate any heterogeneity in the emitter creation process across the flake. These measurements were carried out at 4.5 K, 155 K, and RT, enabling us to examine (i) whether multiple emitter species are generated, (ii) how emitter formation correlates with the spatial location of electron beam exposure, and (iii)  how emitters at individual spatial locations exhibit temperature-dependent spectral behavior. Second, the electron-irradiated hBN with green-yellow emitters was subjected to incremental heat treatments in an inert environment. This approach is motivated by the fact that certain organic compounds can become fluorescent upon heating, while intrinsic defects in hBN gain mobility at temperatures above 800 °C \cite{vogl2019atomic, vogl2018fabrication, chen2023annealing}. Organic molecules are generally sensitive to thermal treatment even at relatively low temperatures \cite{zhao2018single, chen2019dibenzo,macklin1996imaging,qazi2021detection}. Therefore, if organic molecules contribute to the observed green-yellow emission, changes in emitter density or emission characteristics would be expected upon annealing. Finally, the vertical localization of emitter formation was investigated to clarify the potential role of contamination. Previous studies \cite{neumann2023organic} have suggested that contamination-related emitters in hBN are often located at the hBN/substrate interface. To probe this, we performed plasma etching on an electron-irradiated hBN flake containing green–yellow emitters. The plasma progressively removed the hBN layers from the top surface, enabling us to determine the vertical localization of the emitters. A sub-10~nm thick flake was intentionally selected to evaluate whether electron-beam-induced emitter formation is viable in ultrathin hBN.  Based on these experimental results, we then evaluate the hypothesis of a contamination-based origin and, more broadly, contribute to the understanding needed for the reliable implementation of hBN quantum emitters in future quantum technologies.\\

\section{Results and Discussions} 
\begin{figure*}
    \includegraphics[width = 1\textwidth]{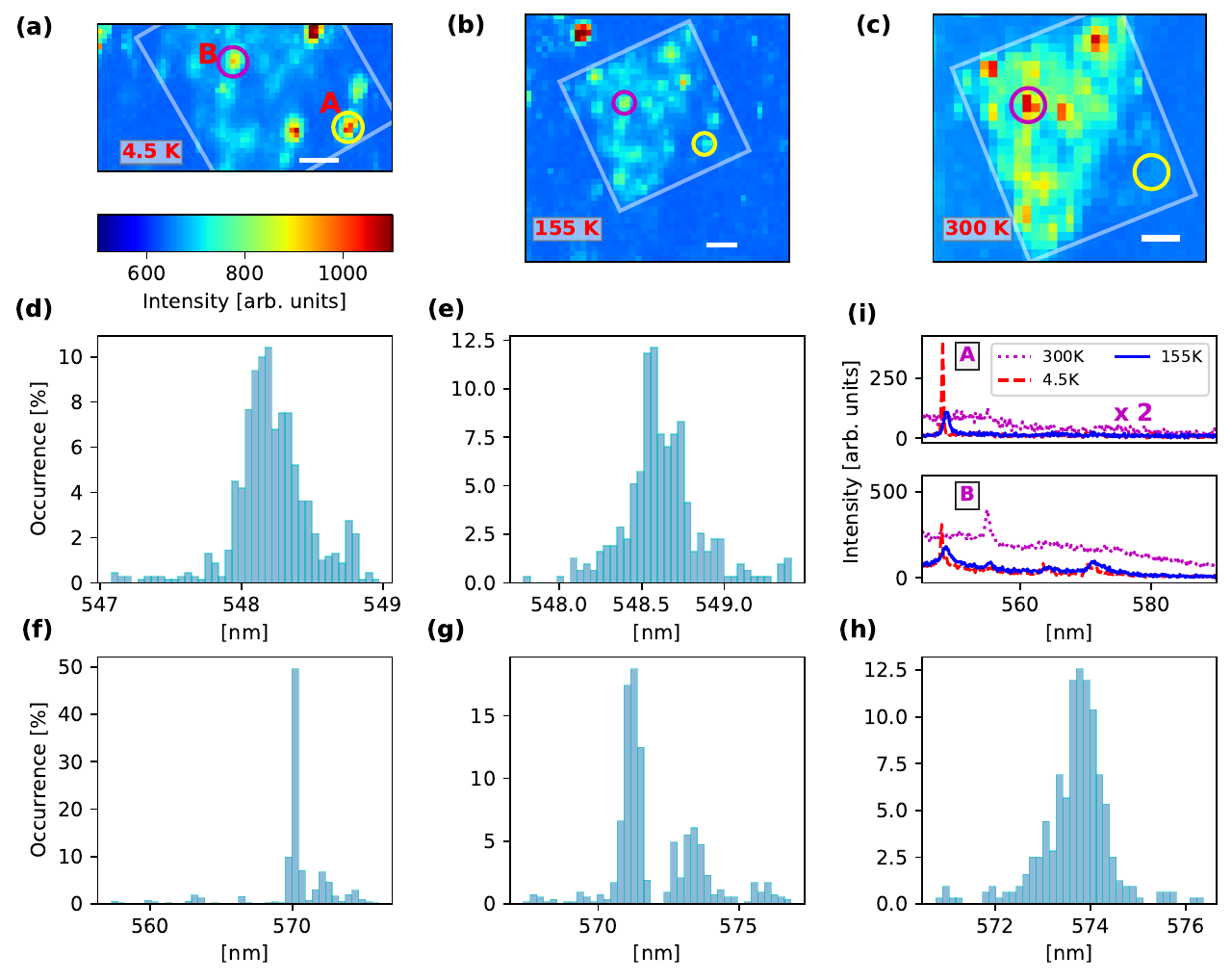}
\captionsetup{
        justification= raggedright, 
        font= small  
    }
    \caption{Hyperspectral images of electron-irradiated hBN . (a--c) Intensity maps of ZPL-filtered emission acquired with an excitation power of 180~$\mu$W at temperatures of 4.5~K, 155~K, and RT, respectively. The irradiated regions are marked by gray squares. The ZPL and PSB at each pixel were fitted using Lorentzian and Gaussian functions, respectively. The extracted peak positions were compiled into histograms for the ZPL (d--e) at 4.5~K and 155~K, and for the PSB (f--h) at 4.5~K, 155~K, and RT. The ZPL peaks are centered at 548.2~nm (4.5~K) and 548.5~nm (155~K), while the PSB peaks are observed at 570.3~nm (4.5~K), 571.3~nm (155~K), and 573.8~nm (RT), indicating a temperature-dependent spectral shift. (i) Representative spectra from two spatial locations indicated in (a). Scale bars in (a--c) correspond to 1~$\mu$m.}
    \label{Figures/HM}
\end{figure*}

\subsection{Hyperspectral Mapping of Electron-Irradiated hBN}
\indent Our approach to assess the spectral variability of emitter creation is raster scanning the flake and recording the spectra. This contains more information than a standard single-photon detector scan, as it provides full spectral information at each spatial position across the region of interest. Later, it can be processed to emphasize individual parts of the spectra over broad spectral backgrounds. This is particularly important for irradiated hBN, where electron-beam exposure simultaneously creates quantum emitters and carbonaceous deposits (see Figure S1-S3 in Supplementary Information). When the full spectral range is integrated into a single intensity map, the resulting image is often dominated by the broadband luminescence from the carbonaceous layers, masking the spatial signature of individual emitters. In contrast, by integrating only a narrow spectral window around the characteristic peak of an emitter, the resulting intensity map reveals diffraction-limited localized emitters, even when they are embedded in regions with strong background emission (see Figures.~S4-S5 in Supplementary Information). A squared area of 5 $\times$ 5 $\mu$m$^2$ was irradiated in order to understand the spatial variability of emission spectra. Atomic force microscopy (AFM) measurements revealed a height increase, and low-energy SEM imaging showed a darker contrast in the irradiated regions (see Figures S1-S3 in the Supplementary Information). These observations are consistent with electron-beam-induced deposition of a carbonaceous film originating from residual hydrocarbons in SEM environments \cite{postek1996approach,nedic2024electron}. Hyperspectral maps at different temperatures zoomed around the ZPL in Figure. \ref{Figures/HM}(a-c) helped us visualize how the emission from spatially localized emitters evolves with temperature. We have performed a statistical analysis of ZPL and phonon sideband (PSB) emission peaks as shown in Figure. \ref{Figures/HM}(d--e) and \ref{Figures/HM}(f-h), respectively. This reveals the presence of a single class of emitter with ZPL centered around 548.2~nm and PSB near 570.3~nm at 4.5 K. With increasing temperature, they got redshifted and broadened (see  Figure. S6-S7 in Supplementary Information), consistent with previous reports \cite{hazra2026temperature} and generally attributed to the narrowing of the bandgap at elevated temperatures \cite{du2014temperature,o1991temperature,tan2022donor}. The ZPL emission variability is seen to be $<$ 1~nm and the PSB peak shapes vary depending on the spatial location. As shown in Figure. \ref{Figures/HM}(i), at some locations the PSB peaks are seen to be split, and their relative height is also 
seen to change. This advanced splitting has been observed in other hBN emitters before \cite{fischer2023combining}. We speculate this as the domination of different phonon modes depending on the local environment. The electron--phonon coupling also varies among emitters, which influences their visibility at elevated temperatures. Emitters with weaker electron--phonon coupling (i.e., a weaker PSB, such as emitter A in Figure.~\ref{Figures/HM}) become difficult to observe, as the RT emission is mostly dominated by the PSB. The ZPL at RT is strongly broadened and partially overlaps with the residual laser background and the hBN Raman (see Figure.~S4 in Supplementary Information), preventing reliable extraction of peak features. For this reason, the RT ZPL was excluded from the statistical analysis. Notably, the emitters maintain high single-photon purity at RT, even when the emission is only collected through PSB (see Figure. S3 in Supplementary Information).  Therefore, we considered only PSB of the green-yellow emitter for further investigations at RT, as evident in the later sections. Across the investigated temperature range, no additional distinct emitter species were observed within the spectral window of 532–600~nm. Overall, hyperspectral mapping validates the creation of a single class of emitters in electron-irradiated hBN, with emitters predominantly localized within the irradiated region. We did not observe large variability of emission energy and spectral shape as typically observed in organic molecules and other reported hBN emitters. 
\subsection{Thermal Stability and Contamination Effects in Electron-Irradiated hBN}
\begin{figure*}
    \includegraphics[width = 1\textwidth]{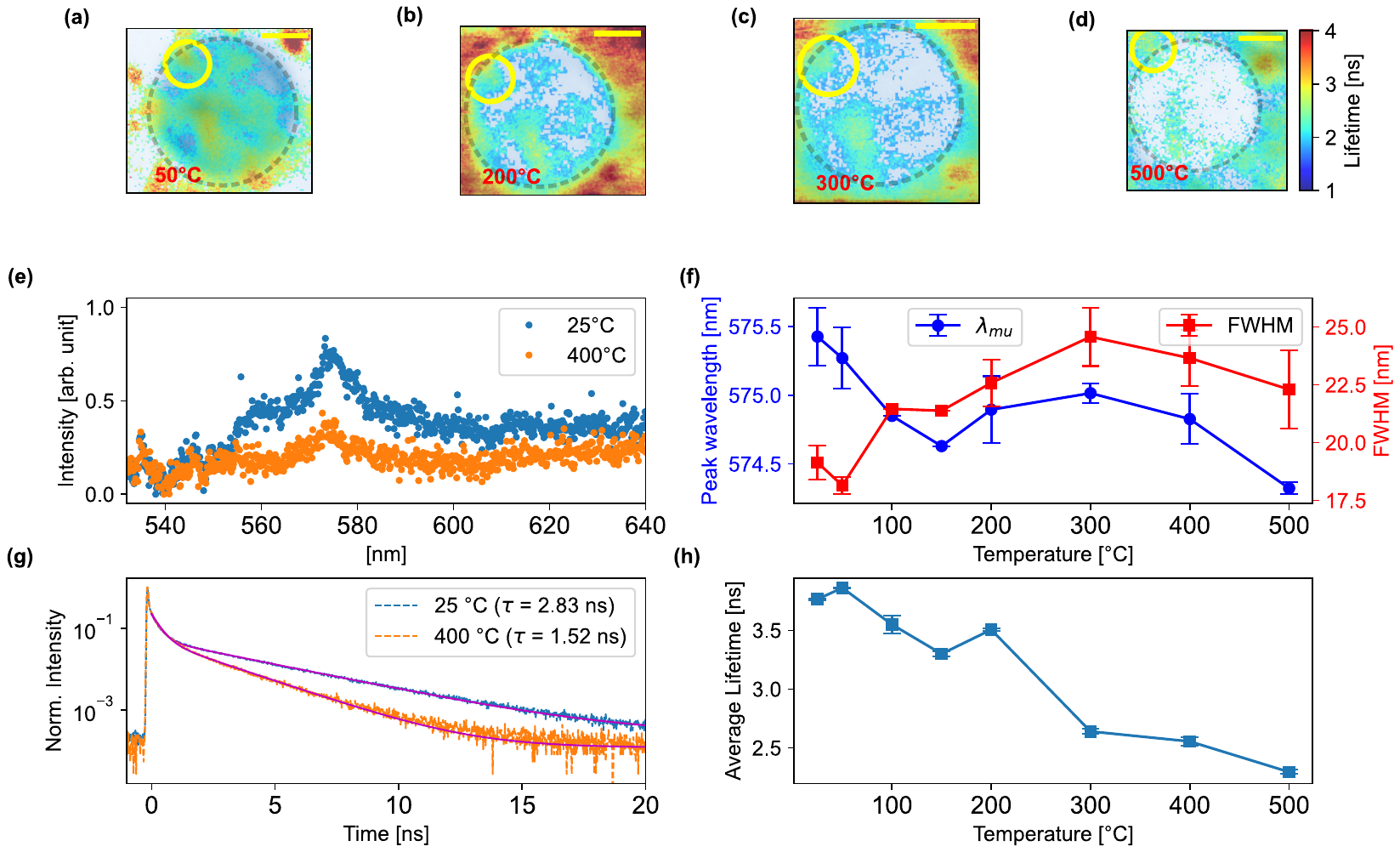}
\captionsetup{
        justification= raggedright, 
        font= small  
    }
    \caption{Photophysical characterization of emitters following 30~min annealing in an Ar atmosphere at different temperatures. (a--d) PL lifetime maps of a irradiated region (marked as grey dotted circle) after annealing at 50 °C, 200 °C, 300 °C, and 500 °C, respectively. These maps reveal the emergence of thermally activated emitters outside the irradiated region and highlight a fundamental difference in emitter density between irradiated and non-irradiated areas. (e) Representative PL spectra recorded at two different annealing temperatures. (f) Central wavelength and FWHM of the PSB as a function of annealing temperature. (g) Lifetime measured at two temperatures. (h) Average lifetime as a function of temperature. The measurements shown in (e) and (g) correspond to the emitter highlighted by the yellow circle in panels (a–d). Scale bars in (a--d) correspond to 1~$\mu$m.}
    \label{Figures/annealing}
\end{figure*} 
\indent In addition to the broad spectral variability, certain organic compounds are known to exhibit fluorescence upon thermal treatment \cite{neumann2023organic,neumann2019accidental,chen2019dibenzo}. Annealing under an inert environment, therefore, serves as a tool to investigate the presence of organic contamination in the emitter creation process. We annealed the electron-irradiated hBN containing the green-yellow emitters in an Ar environment in incremental temperature steps. The photophysical characteristics of the emitters were recorded for each annealing temperature after the sample cooled down to  RT.  This probes the thermal stability of green-yellow emitters and checks if there is contamination present in the sample that gets thermally activated. The photoluminescence (PL) lifetime maps of a 3 $\mu$m-wide circular irradiated region after annealing at different temperatures are shown in Figure~\ref{Figures/annealing}(a--d). At 50~°C, several green-yellow emitters with PL lifetimes exceeding 3~ns are observed within the irradiated region. One representative emitter is highlighted by the yellow circle in the maps. To improve the visibility of the long-lifetime emitters, pixels with lifetimes above 1~ns are displayed in the lifetime maps. The corresponding PL intensity maps are provided in Figure.~S8 of the Supporting Information. At 50°C there is hardly any fluorescence outside the irradiated region. Upon annealing above 150~°C, a high density of thermally activated emitters emerges uniformly throughout the non-irradiated regions. But their density remains strongly suppressed within the irradiated area. These thermally activated emitters have a relatively higher lifetime compared to the green-yellow emitter, as evident in Figure.~\ref{Figures/annealing}(b-c). This behavior highlights a fundamental difference between emissions from irradiated and non-irradiated regions.   Further distinction arises from their photophysical properties: emission from the non-irradiated areas is broad, undergoes photobleaching under prolonged laser excitation and fades away in high-temperature annealing (see Figure. S9 in Supplementary Information).  This indicates a contamination-related origin \cite{qazi2021detection}. In contrast, the green-yellow emitters localized within irradiated regions exhibit stable emission without blinking or bleaching, as shown in Figure. S10 Supplementary Information. Their emission spectra (Figure. \ref{Figures/annealing} (e–-f)) remain unchanged in the range of observed annealing temperatures up to 500 °C. Usually at this temperature range, atomic reconstruction is improbable \cite{vogl2019atomic}. An unchanged spectrum also indicates that the underlying defect structure of the emitter remains unchanged.  Moreover, the persistent suppression of contamination-related fluorescence in irradiated regions suggests that electron irradiation alters the local environment in a way that prevents the activation of unwanted emitters. Notably, a gradual decrease in total lifetime and emission intensity was observed as shown in Figure. \ref{Figures/annealing} (g–-h) and Figure. S10 in Supplementary Information, respectively. This may arise from the activation of additional nonradiative decay channels upon annealing. Additionally, both spectral and lifetime measurements across the flake remain comparatively homogeneous at each annealing step, without evidence of strongly varying sub-populations that might be expected from a distribution of independent contaminants with different thermal stabilities. The slight red-shift in peak energy ($\sim$ 1~nm) may be associated with a change in local environment or strain modification in the sample, although this requires further investigation. The observed narrowing of the full width at half maximum (FWHM) after high-temperature annealing can be a fitting artifact. As the Gaussian fit may capture only the clearly resolved part of the spectrum, leading to an apparent narrowing. Alternatively, the narrowing may result from reduced inhomogeneous broadening due to partial cleaning during annealing.\\
\indent The emitters are no longer detectable after annealing above 500 °C. The irradiated regions appear significantly darker (see  Figure. S10 in Supplementary Information), consistent with the removal of carbonaceous deposits and the associated suppression of broad background emission \cite{kumar2023localized,nedic2024electron}. The loss of detectable signal from the emitters is indicative of a substantial reduction in quantum efficiency, reflected in the gradual decrease in both lifetime and intensity with increasing annealing temperature. This is likely caused by changes in the local defect environment due to high-temperature annealing. Notably, diffraction-limited green–yellow emitter spots in the irradiated regions remain preserved up to 500 °C, as highlighted by the yellow circles in PL maps in Figure.~\ref{Figures/annealing}(a-d) and S8 in Supplementary Information. This suggests that the emitters do not disappear progressively with annealing. If contamination were dominant, a broader distribution of thermal responses might be expected. Instead, emitters across the flake exhibit relatively homogeneous behavior in both spectral signature and lifetime, as well as in the eventual loss of optical signal. \\
\indent Overall, these observations suggest that the defect structure remains intact with annealing up to 500 °C, while the recombination dynamics are significantly affected. Together, our observations indicate that our samples are not free from contamination. But there is a clear distinction between thermally activated contamination fluorescence and the robust green-yellow emitters, supporting a non-contamination-related origin for the latter. Besides, it also highlights an additional benefit of electron-beam irradiation in suppressing unwanted background emission. 
\subsection{Vertical Localization of the Green-Yellow Emitters}

\begin{figure*}[ht!]
    \centering
    \includegraphics[width=1\textwidth]{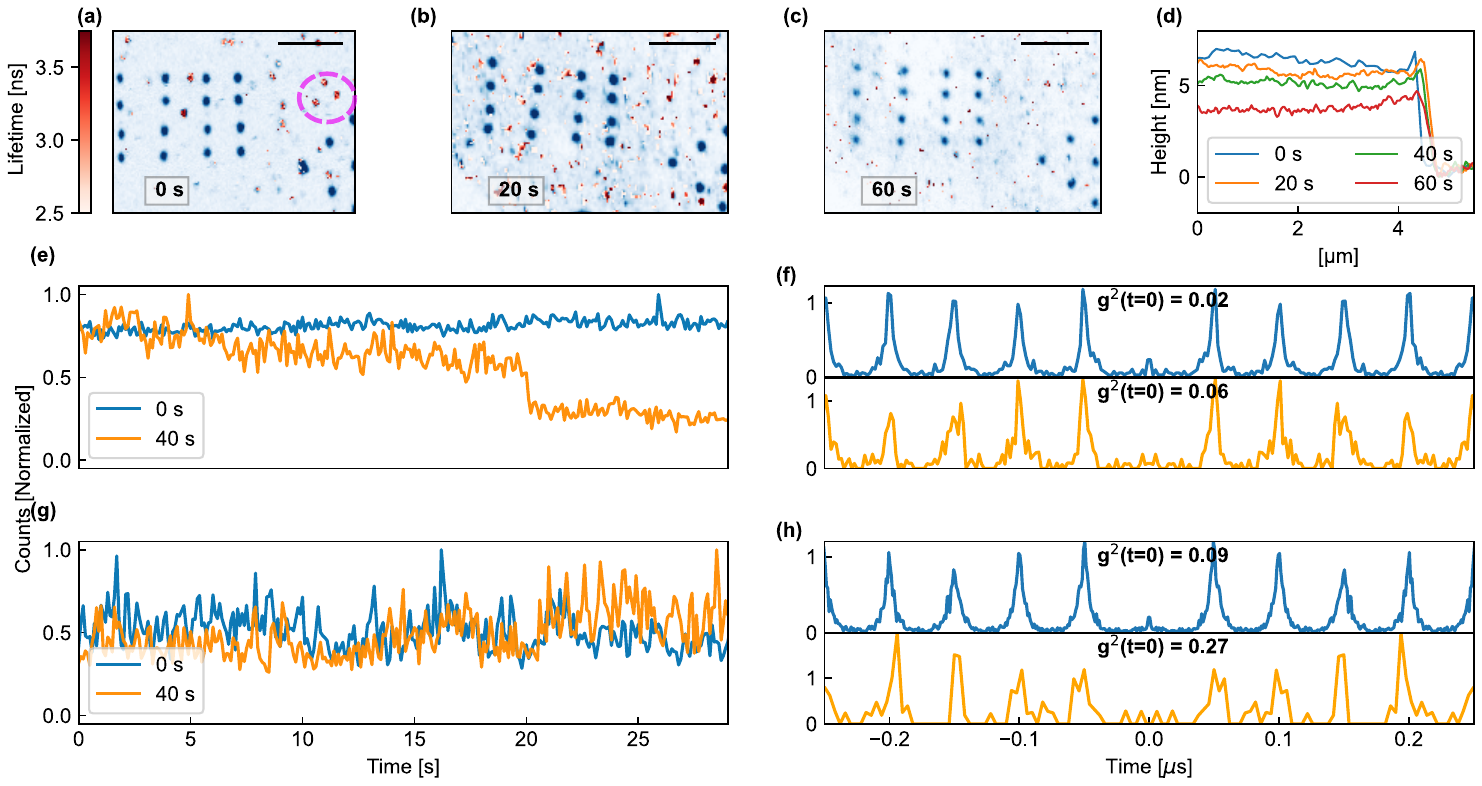}
    \captionsetup{
        justification=raggedright
    }
\caption{Layer-by-layer etching of an hBN flake. PL lifetime maps recorded (a) before, (b) after 20 s, and (c) after 60 s of oxygen plasma etching. Blue dots indicate the irradiated spots on the hBN flake, while red dots mark the emitters with 3-5 ns lifetime. The pink circle in (a) highlights emitters created far from the irradiated region. (d) hBN thickness function of plasma etching time. (e) PL intensity time trace of an emitter exhibiting photobleaching after 40 s of plasma etching, with the corresponding second-order autocorrelation function, (g$^{(2)}(\tau)$), shown in (f) before (blue) and after (orange) etching. (g) PL intensity time trace of an emitter located in the thinnest region of the flake, confirming emitter creation in an hBN flake with an initial thickness of 6.4~nm. The emitter remains photostable after 40 s of plasma etching, corresponding to a reduced flake thickness of 5.2~nm. The corresponding (g$^{(2)}(\tau)$) measurements before (blue) and after (orange) etching are shown in (h). The scale bar in (a)–(c) is 5 $\mu$m.}
    \label{Figures/z_loc}
\end{figure*}
\indent Beyond photophysical signatures, spatial localization provides an additional route to distinguish intrinsic emitters from contamination-related origins. Hyperspectral mapping reveals that the green-yellow emitters are deterministically generated at the electron-beam irradiation sites. However, while this lateral correlation confirms their association with the irradiated regions, it does not reveal their vertical position within the hBN flake. Previous studies have suggested that contamination-related emitters preferentially form at the hBN/substrate interface \cite{neumann2023organic}. This motivates a direct investigation of the vertical localization of the emitters to determine whether they are located within the hBN crystal or at the hBN/substrate interface.\\
\indent To probe this, the irradiated hBN flake with green-yellow emitters was thinned progressively using oxygen plasma while tracking the number of emitters after each etching step. PL lifetime maps depicted in Figure.~\ref{Figures/z_loc}(a–c) give an overview of the emitter population as thinning of the flake occurs. The blue dots indicate the irradiated spots, and the red dots indicate emitters that have a lifetime of 3--5 ns. The thinning as a function of etching time is depicted in the AFM line profile in Figure.~\ref {Figures/z_loc}(d). Before any plasma etching, the green-yellow emitters were identified and characterized. After the first 20 s etch, several new emitters appeared, visible as red spots in the lifetime map in Figure.~\ref{Figures/z_loc}(b). This is expected, as oxygen plasma itself can activate new luminescent defects in hBN \cite{vogl2018fabrication}. Nevertheless, correlating the post-etch confocal PL maps with the initial maps enabled us to identify the emitters that existed prior to plasma treatment and distinguish them from plasma-induced emitters (as shown in Figure. S11-S14 in the Supplementary Information). After an additional 20 s of etching, some of the green-yellow emitters became unstable and photobleached during optical measurements, as shown in Figure. 3(e–f). This behavior indicates that the emitters were not yet physically removed; rather, removal of the upper hBN layers strongly affected their stability. Similar behavior has been reported for other hBN quantum emitters \cite{liu2025single, yamamura2025quantum}, as emitter stability strongly depends on the surrounding crystal environment. \\
\indent In Figure.~\ref{Figures/z_loc}(g), we demonstrate the time-resolved PL counts from a green-yellow emitter in a hBN flake of thickness 6.4~nm. When it was further etched to 5.2~nm, we still observed emission from the same emitter with single-photon purity (Figure.~\ref{Figures/z_loc}(h)), although the emission was weaker. The slightly higher (g$^{(2)}(0)$) value after etching is due to the reduced signal-to-noise ratio. Similar behavior for electron-irradiated emitters (B-centers) in very thin hBN has been reported previously \cite{liu2025single}. The creation of emitters in such thin flakes has been correlated with electron scattering, where emitters can be found several microns away from the irradiated spots. We also observed emitters far from the irradiated spots, as depicted by the pink circle in Figure. \ref{Figures/z_loc}(a) and Figure. S11 in the supplementary information, although this requires further investigation.\\
If the emitters were located at the hBN/substrate interface, they would be expected to persist irrespective of the hBN flake thickness. But we did not observe any green-yellow emitters in flakes thinner than 5.2~nm. The irradiated regions still show bright emission in the PL map, but no green-yellow emitters are found at this thickness, as shown in Figure. S13 in the supplementary information. In the thicker parts of the flake, emitters are still present upon further thinning, as shown in Figure. S11-S12 of the Supplementary Information. Taken together, these results argue against an origin exclusively at the hBN/substrate interface. Instead, they show that the green-yellow emitters are distributed within the hBN crystal while exhibiting a dependence on the surrounding environment.

\subsection{Discussion}
A simplified emitter fabrication process that omits cleaning steps introduces the risk of extrinsic contamination. Since many organic molecules can act as efficient visible single-photon emitters, it is essential to carefully assess a possible extrinsic origin. In the experiments described above, we therefore compare key properties of organic emitters with those of green–yellow hBN emitters, namely spectral variability, response to thermal treatment, and spatial as well as vertical localization. The strong spatial correlation with electron beam irradiation suggests that any contamination-related mechanism would also need to be beam-induced. However, previous studies \cite{kumar2024comparative} have shown that similar irradiation conditions do not produce comparable emitters in other wide bandgap materials such as GaN, SiC, and mica, arguing against a generic contamination pathway. Furthermore, hyperspectral mapping reveals a well-defined and narrow spectral distribution, in contrast to the broad and heterogeneous emission typically associated with contamination \cite{neumann2019accidental}. We also note the absence of spectral wandering or broad distributions of emission energy that are commonly observed for organic emitters. Spatially, the emitters are localized around the irradiated regions on the flake rather than at the flake edge, where organic molecules were reported previously in hBN \cite{neumann2023organic}. In addition, plasma etching experiments reveal the persistence of emitters even after removal of several hBN layers from the top that is difficult to reconcile with a purely surface-bound contaminant. \\
\indent Annealing experiments provide further insight into the nature of these emitters. Both spectral and lifetime measurements across the flake remain homogeneous at each annealing step, with no evidence of a varying density of diffraction-limited spots in the irradiated region up to 500 °C. This is not necessarily expected for a distribution of independent contaminants with different thermal stabilities and binding environments. The loss of optical signal from the green–yellow emitters observed after annealing above 500 °C can, in principle, arise from several mechanisms; here, we speculate on two plausible scenarios. First, the emitters could be associated with interstitial-type defects, which are expected to become mobile at comparatively lower temperatures than substitutional or vacancy defects, given that intrinsic lattice defects in hBN generally require higher temperatures ($\sim$800 °C) for migration \cite{vogl2019atomic, zobelli2007vacancy}. This can be further supported by the fact that the PSB appears at $\sim$90 meV from the ZPL \cite{hazra2026temperature}, which is lower than the typical longitudinal optical (LO) phonon energy in hBN \cite{serrano2007vibrational}. On the other hand, the out-of-plane optical (ZO) phonon mode has been reported at similar energies in interstitial defects \cite{hoese2020mechanical} . One can also expect changes in emitter characteristics with increasing annealing temperature in such a case. Although a $\sim$1~nm shift in the PSB spectrum is observed at RT, this can also arise from spectral diffusion or changes in the local dielectric or strain environment. Alternatively, the emitters may correspond to in-plane defects whose radiative efficiency is strongly influenced by their local environment. In this picture, annealing above 500 °C may influence the SEM-induced carbonaceous layer and/or other surface adsorbates, thereby modifying the local dielectric and charge environment. This can enhance non-radiative decay channels, resulting in a reduction in emission intensity and lifetime. The emitters exhibit a dominant PSB compared to ZPL at RT, emphasizing strong coupling to their surroundings. Additionally, the emitters are located several layers beneath the surface, suggesting that they cannot be attributed purely to surface defects. Taken together, these observations argue against a purely contamination-based origin of the emitters. Instead, the results support a model in which intrinsic or lattice-bound defects are responsible for the emission, while their optical activity is strongly modulated by the local environment.

\section{Conclusion}
\indent In summary, a combination of experiments reveals several key characteristics of the green–yellow emitters in hBN that provide insight into their microscopic origin. The emitters are densely generated in regions exposed to electron-beam irradiation. Their creation is demonstrated even in flakes as thin as ~6.4~nm, and stable emission from SPE is still observed after further thinning to 5.2~nm, highlighting their compatibility with nanoscale and near-field photonic platforms where reduced thickness is advantageous. SEM and AFM imaging indicate local modification in the irradiated regions, indicating electron-beam-induced deposition of a carbonaceous layer related to hydrocarbon contamination. Hyperspectral mapping further shows a strong spatial correlation between these modified regions and the emitter distribution, suggesting a possible role of such residues or carbon-related species in the formation process. At RT, the emission is dominated by the PSB, which indicates a strong coupling to the environment. Annealing experiments provide additional insight into their stability and environmental sensitivity. The emitters exhibit stable spectral signatures without blinking or bleaching up to 500 °C, indicating a robust underlying defect structure. The pronounced sensitivity of the emission intensity and lifetime to thermal treatment highlights the strong coupling between these emitters and their local environment. This environmental dependence suggests potential for sensing applications in integrated photonic systems where local heating may play a role. Overall, the experimental evidence rules out a purely surface adsorbates/organic molecular contamination-based origin for the observed green–yellow quantum emitters in hBN. However, the possibility of interstitial contributions, or a dependence on the SEM-induced carbonaceous deposition, cannot be fully excluded and needs further investigation. Future experiments could help disentangle these effects, for example, by applying in-plane and out-of-plane electric fields to probe dipole orientation \cite{zhigulin2023stark,nikolay2019very,noh2018stark} and/or by systematically controlling and reducing hydrocarbon exposure in the SEM chamber \cite{nedic2024electron} to directly assess its role in emitter activation and quenching.

\section{Experimental Section}
\threesubsection{Emitter Creation }\indent All the emitters in this work were created by electron irradiation in exfoliated hBN flakes, as described in previous work \cite{kumar2023localized}. A specific area was defined in the software and filled with electron beam irradiation with  7.7 $\times$ 10$^{17}$ cm$^{-2}$ fluence, 3 kV acceleration voltage, 25 pA current, and 10 seconds dwell time. No pre- or post-treatment was required. \\

\threesubsection{Sample Treatment }\indent To investigate the thermal stability of the emitters, annealing was performed using a tube furnace system (AccuThermo AW-610). Prior to heating, the chamber was purged with nitrogen (N$_2$) to remove residual oxygen. The samples were then annealed under a flowing argon (Ar) atmosphere at the target temperature for specified durations of 30~min. After annealing, the samples were allowed to cool down to RT under Ar environment.\\
\vspace{5pt}
\indent For the vertical localization experiment, the hBN layers were etched using a low-power RF oxygen plasma system (SPI Plasma Prep™ plasma etcher/cleaner, Structure Probe, Inc.). The etching was carried out in multiple sequential rounds. In the first three rounds, the sample was exposed to plasma for 20 s each, at chamber pressures of approximately 575 mTorr, 620 mTorr, and 700 mTorr, respectively. For all etching steps, the RF power was maintained at 100 W.\\
\threesubsection{Optical Characterization}\indent Hyperspectral imaging measurements were performed using a closed-loop cryostat (attoDry 800, attocube systems AG). The sample was excited with a 515~nm diode laser (Roithner Lasertechnik GmbH). The excitation beam was passed through a linear polarizer followed by a quarter-wave plate to generate circularly polarized light, ensuring uniform excitation of emitters regardless of their dipole orientation. The beam was then focused onto the sample using a low-temperature compatible objective with an NA of 0.82 (attocube systems AG). The PL signal from the sample was collected through the same objective in a confocal configuration. The collected emission was spectrally filtered using a 532~nm long-pass filter to suppress the excitation laser and subsequently directed to a spectrometer (Teledyne Princeton Instruments) for spectral acquisition.\\
\indent For temperature-dependent measurements, the sample temperature was controlled by varying the current through a 100 $\ohm$ carbon resistor placed below the sample. Temperature calibration was performed using a 1 k$\ohm$ Allen-Bradley carbon resistor. \\
\indent Photophysical characterization following oxygen plasma treatment and annealing was performed at RT. At RT, the analysis was based on the PSB because the green-yellow emitters exhibit a well-resolved PSB while maintaining single-photon purity \cite{kumar2023localized,kumar2024polarization}. The emitters were excited using a 532~nm, 80 MHz laser coupled to a commercially available confocal PL setup (MicroTime 200, PicoQuant). The 532~nm excitation wavelength was selected based on our previous finding that excitation through the PSB results in a significantly enhanced emission intensity \cite{hazra2026temperature}. The excitation beam was circularly polarized to ensure uniform excitation of emitters with arbitrary dipole orientations. The emitted PL was separated from the excitation path using a 550~nm dichroic mirror. For measurements following oxygen plasma treatment, an additional 550~nm long-pass filter was used in the detection path. For annealed samples, a 532~nm long-pass filter was employed to allow broader spectra collection. Due to the fixed 550~nm dichroic mirror in the optical path, spectral information below 550~nm is  attenuated and therefore not considered reliable in the analysis.

\medskip
\textbf{Supporting Information} \par 
Supporting Information is available from the Wiley Online Library or from the author.

\medskip
\textbf{Acknowledgements} \par  This research is part of the Munich Quantum Valley, which is supported by the Bavarian state government with funds from the Hightech Agenda Bayern Plus. This work was funded by the Deutsche Forschungsgemeinschaft (DFG, German Research Foundation) under Germany’s Excellence Strategy - EXC-2111-390814868 (MCQST), under Project-ID 398816777 - SFB 1375 (subproject C2) and the International Research Training Group (IRTG) 2675 Meta-Active (project number 437527638). The authors acknowledge support from the Federal Ministry of Research, Technology and Space (BMFTR) under grant number 13N16292 (ATOMIQS) and by the German Space Agency DLR with funds provided by the Federal Ministry for Economic Affairs and Climate Action (BMWK) under grant numbers 50WM2165 (QUICK3) and 50RP2200 (QuVeKS).

\medskip
\textbf{Notes} \par 
We acknowledge the use of AI assistance to improve the Python script that plotted the figures and rephrased parts of the manuscript for readability. All scientific interpretations, data analyses, and final writing decisions were made by the authors. The authors declare no competing  financial interest. 

\medskip

%

\bibliographystyle{MSP}
\bibliography{main}




\end{document}